\documentclass[a4paper,twoside,twocolumn]{article} 
\usepackage{graphicx}  
\usepackage{graphics} 
\usepackage{fancyhdr}
\newcommand{\dfr}{d\raise0.3ex\hbox{\kern-0.5ex\char"013 }}  
\newcommand{\lapla}{\bigtriangleup} 

\begin{document}

\title{\textbf{On the morphogenesis of stellar flows}\\ \it{Application to planetary nebulae}} 
\author{D. Da Rocha$^{1}$ and L. Nottale$^{1}$\\ 
\footnotesize{$^{1}$UMR CNRS 8631, LUTH, Observatoire de Paris-Meudon, F-92195 Meudon Cedex, France} }
\date{\footnotesize{in original form 2003 April 14}}

\maketitle

\begin{abstract} A large class of stellar systems (e.g., planetary nebulae (PNe), supernova envelopes, LBV
stars, young stars in formation) shows structures in their accretion/ejection phase that have similar
characteristics. In particular, one currently observes for these objects equatorial discs, axial ejections
and stable bipolar shells. However these simple shapes, which are expected to be solutions of standard
hydrodynamical equations, are not yet fully understood. In this paper, we suggest a new form of these
equations that takes into account the fractality and the irreversibility of particle motion in such
processes. Then we study in this framework a general infall or ejection motion in a central spherical
potential. From the stationary solutions allowed by this new hydrodynamical system, we deduce a specific
distribution of matter density, described in terms of probability density for ejection angles. A global
classification of predicted shapes, depending on the values of conservative quantities such as
($E^2,L^2,L_z$), is given. These results are compared with the available observational data, and allows us
to theoretically predict the possible existence of more complicated structures and of correlations between
observable variables, which could be checked by future observations. \end{abstract}
\\
\\
\textbf{keywords}: relativity - gravitation - hydrodynamic - planetary nebulae - LBV star - supernovae - star
formation

\section{Introduction} With the recent evolution of
observational techniques, a great diversity of unexpected gravitational structures has spring up in many
astrophysical domains. The large number and the stability of these structures has pointed out the
incompleteness of their understanding in terms of standard approaches. This still represents a large and
fundamental open problem. Moreover, one observes similar structures in many different systems, at many
different scales and in various conditions characterizing the underlying medium, such as stable discs,
rings, disc/jets combinations and bipolar outflows.\\ The present works on the formation and the dynamics of
planetary nebulae are based on hydrodynamical collisions between slow and fast winds emitted at different
stellar evolution stages. This is the Interacting Stellar Wind (ISW) model, generalized by Balick \cite{balick}.
In this model, slow winds (10-20 km/s) are firstly emitted during the AGB phase then, at the end of this
period, UV-radiations ionize this circumstellar slow medium. After this initial ejection, the stellar
surface still losses its matter, but fast winds ($\approx 1000\,$ km/s) are observed and enter in collision
with slow winds. This collision induces a compressed zone where a hot bubble creates the observable bright
shells of planetary nebulae. The spherical and elliptic shapes are satisfactorily understood in the
framework of this description (\cite{dyson}, \cite{kwok}), but the majority of the planetary nebula
shapes cannot result from this model \cite{franco}. Despite the improvement of hydrodynamic simulations,
even the simplest bipolar shells are not found to be formed in an universal way \cite{balick2}, since each
particular case needs an ad-hoc hypothesis on the density of presumed slow equatorial winds and observed
fast winds. Moreover, bright bipolar shells are not directly connected to the hot bubble resulting from the
collision \cite{balick}. Thus, in the standard hydrodynamical approach, the initial hypothesis made about
the nature of the slow and fast winds (which are connected to each central system studied) play a central
role.\\ 
In the present paper, we shall explore a different though complementary approach to these problems.
Independently of the initial wind densities, we can study the consequences of the specificity of individual
trajectories of particles making the shells.  We use a fluid-like description, in which the possible
trajectories are non-deterministic. Because of the information loss due to collisions and to the continual
ionization/diffusion of central star photons, individual particle trajectories can no longer be strictly
followed, so that a statistical description in terms of fractal and non-differentiable trajectories is
adopted. Such a description finally gives a new form to the hydrodynamical equations. 

\section{Theory} In the following
theoretical development, we use general hydrodynamical concepts, then we apply them to the specific case of
an ejection process. It can be already noted that these solutions can be also used, with a mere time
reversal, for describing infall motion in a spherical potential.\\ In a typical process of stellar mass
ejection (developed above), the interaction between winds or ISM and the continual interaction of emitted
photons allow one to set the following simple hypotheses:

 (i) For each particle, the information about its trajectory is lost beyond some small space-scale and
time-scale (compared with the global space-scale and time-scale of structuration of the system). Such a
memory loss of the previous dynamical conditions is compatible with a fractal description using a fractal
dimension $D_F = 2$ for trajectories (as in standard Brownian motion). Only global informations are
conserved. 

(ii) The test particles can follow an infinity of potential trajectories, and the study of such systems
therefore comes under a fluid approach (like in their standard description). 

(iii) The last main point is the irreversibility of this process. This irreversibility is local and global:
for small scales, this is connected with the non-deterministic trajectories, and at the scale of the global
system, this represents the impossibility for the system to come back to its initial conditions.\\ These
three points can be summed up in terms of three fundamental conditions which lead to the construction of new
hydrodynamic tools: 

(i) The fractality of each individual trajectory beyond some transition scale. It is described in terms of
elementary displacements which read: 
\begin{equation} dX=dx + d \xi, \end{equation} 
where $dx=v dt$ is a classical, differentiable variable, while $d\xi$ is a non-differentiable, stochastic variable, that
describes the fractal fluctuation. Its non-deterministic character implies that it is known only through a
statistical description. Namely, in the simplest case only considered here (fractal dimension 2), it is such
that \begin{equation} <d \xi>=0,   \;\;\;   {\rm and}   \;\;\; <d \xi^2>= 2 {\cal D} dt. \end{equation} Its
non-differentiability is apparent here since $<d \xi^2>^{1/2}/dt \propto dt^{-1/2}$ and it is therefore
formally infinite. However, as we shall see, the fact that it vanishes in the mean allows one to give a
description in terms of only the differentiable part $dx$ of the elementary displacement, that includes the
indirect effects involved by the existence of the non-differentiable part. The coefficient ${\cal D}$ is a
measure of the amplitude of the fractal fluctuations.

(ii) The test-particles can follow an infinity of possible trajectories: this leads one to jump to a
non-deterministic, fluid-like description, in terms of the ``classical part" of the velocity field of the
family of trajectories, $v=v(x(t),t)$. 

(iii) The reflection invariance under the transformation ($dt \leftrightarrow -dt$) is broken as a
consequence of the non-differentiability. Indeed, there are two definitions of velocity which are based on
the variation of the position variable $X$ either in the interval $[t-dt,t]$, or in the interval $[t,
t+dt]$. The two definitions are equivalent in the differentiable case, but no longer in the
non-differentiable one considered here. This leads to a two-valuedness of the velocity vector, $(v_-,v_+)$.
The use of a complex velocity, ${\cal V}= ({v_+}+{v_-})/2 - i ({v_+}-{v_-})/2$ to deal with this
two-valuedness can be shown to be a covariant and simplifying representation ( Nottale \cite{liwos}, C\'el\'erier 
\cite{dirac}). 

These three effects can be combined to construct a complex time derivative operator (\cite{liwos},
\cite{Chaos1}, \cite{quantuniv1}), that reads,  
\begin{equation} 
\label{complexdif} 
\frac{\dfr}{dt} = \frac{ \partial }{ \partial t } + {\cal V} . {\nabla} - i{\cal D}{\lapla} 
\end{equation}

Now, we can use this new relation and follow the standard classical mechanics construction. A general
Lagrange function ${L}(x,{\cal V},t)$ characterizes any system and we define a general action $\cal S$
thanks to \begin{equation} {\cal S} = \int_{t_{1}}^{t_{2}}{L}(x,{\cal V},t) \,dt. \end{equation} Since the
velocity is now complex, the same is true of the Lagrange function and therefore of the action. The
generalization to a complex velocity $\cal V$ does not modify the form of the classical Euler-Lagrange
equation (\cite{Chaos1}, \cite{quantuniv1}), but now, the classical differential operator $d/dt$ is
replaced by the complex differential operator $\dfr / dt$ \begin{equation} \frac{\dfr}{dt} \, \frac{\partial
L}{\partial {\cal V}_{i}} - \frac{\partial L}{\partial x_{i}}= 0. \end{equation} Let $\mu$ be the mass of
the particle. In the case of a Newtonian closed system in a scalar field which is considered here, the
Lagrange function becomes $L = \frac{1}{2} {\mu} {\cal V}^{2} - \Phi$. This leads to a motion equation that
conserves the form of Newton's fundamental equation of dynamics \begin{equation} {\mu} \frac{\dfr}{dt} {\cal
V} = - \nabla \Phi, \end{equation} which is now written with complex time derivative operator and complex
velocity. The complex velocity $\cal V$ still depends on the complex action according to the standard form
(${\mu} \cal V = \nabla \cal S$). Let us now make a change of variable, and define the function:
\begin{equation} f = \exp \left( \frac{ i \cal S}{2 {\mu} \cal D}  \right) .  \end{equation} In terms of
this function, the fundamental equation of dynamics now writes \begin{equation} 2i{\mu}{\cal
D}\frac{\dfr}{dt} (\nabla \, \ln \,f) = \nabla \Phi. \end{equation} By replacing in this equation the
complex derivative operator by its expression (Eq. \ref{complexdif}), one finds that it becomes
\begin{equation} \label{schroderiv} -2{\cal D}\nabla \left(i \frac{\partial}{\partial t} \ln \, f + {\cal
D}\frac{ \lapla f}{f} \right) = -\frac {\nabla \Phi}{{\mu}}. \end{equation}

The real and the imaginary part of this equation can be separated. Indeed, the complex action is rewritten
${\cal S} = S + i S^{'}$, and so, the function ($f$) takes another form: \begin{equation} f = \exp{
\left(\frac{ i ( S+iS^{'})}{2 {\mu} \cal D} \right) }= \sqrt{\rho} \, \exp{ \left(\frac{ i S}{2 {\mu} \cal
D} \right) }, \end{equation} with $S$ the classical action which is connected to the classical velocity:
${\mu} V = \nabla S$. Finally, the real and imaginary part of the motion equation induce the following
system of equations \begin{equation} \label{AA1} {\mu} \, (\frac{\partial}{\partial t} + V \cdot \nabla) V 
= -\nabla (\phi+Q), \end{equation}

\begin{equation} \label{AA2} \frac{\partial \rho}{\partial t} + {\rm div}(\rho V) = 0. \end{equation} The
first one is a standard Euler-Newton equation but with the appearance of an additional potential energy term
$Q$ that writes: \begin{equation} Q = -2{\mu} {\cal D}^2  \frac{\lapla \sqrt{\rho}} {\sqrt{\rho}} , 
\label{Q} \end{equation} and which is a manifestation of the fractal geometry.  The second one is the
continuity equation. We have therefore recovered the current minimal equations which are used in the
standard hydrodynamical description, except for the new potential term. In the limit where the fractal
fluctuations vanish, i.e. $ {\cal D} \rightarrow 0$, the potential energy $Q$ vanishes and the description
is exactly reduced to the standard one. But conversely, the introduction of this new potential energy allows
one to integrate the hydrodynamics equation under the form of Eq. \ref{schroderiv}, which has the advantage
to be linear and to allow the obtaining of analytical solutions. 

The generalized hydrodynamical system obtained above can now be used as motion equation for a large class of
systems \cite{darocha}, namely, all those coming under the three conditions that underlie its
demonstration: large number of possible trajectories, fractal dimension 2 of trajectories, and local
irreversibility. Actually these conditions amount to a loss of information about angles, position and time.
But, paradoxically, such a total loss of information on the individual trajectories results in a tendency
for self organization and structuration of the systems \cite{quantuniv1}.

\section{Theoretical development: spherical potentials}

\subsection{Symmetry constraints} The resolution of physical problems depends on the conservation laws
imposed by the symmetries to which the physical system under consideration is subjected. As a first step, we
shall consider spherically symmetric systems. Then, for a more complete study, other symmetries will be
explored, in particular axial symmetrical systems. Moreover, the elementary approximation about the
spherical potentials will be enriched by taking into account dynamical perturbations. These various effects
are accounted for in a simple way by a specific choice of  the coordinate systems (e.g. elliptic, parabolic
or cylindrical).

Let us consider a test particle ejected by a star, submitted to various forces described by potentials
(${{\varphi}_i}$). Its motion is described by the hydrodynamical equation Eq.\ref{AA1}:

\begin{equation} {\mu} \, (\frac{\partial}{\partial t} + V \cdot \nabla) V  = -\nabla (\sum{{\varphi}_i}
+Q), \end{equation} Such a system is expected to be subjected to gravitational, radiative pressure,
electromagnetic, and collisional forces.

However, in order to simplify the description we shall base ourselves on a fundamental observational result
concerning these objects: the observations of the shells of many PNe's have shown that the expansion
velocity is nearly constant. Such a result is already used in many numerical simulations (\cite{corra},
\cite{dwar}, \cite{garcia2}). This means that, in the first order approximation (neglecting the
interaction with the ISM), there is a dynamical compensation between the various forces acting on the shell,
in particular between the gravitational force and the radiative stellar pressure. We shall therefore assume,
as a first simplifying step, that the potential is almost constant. We take the constant velocity
approximation (i.e. free test particles) as a starting point for the following developments: In a second
step, the model will include elements of perturbation such as velocity expansion to second order and the
first order of perturbation power-series expansion (magnetics contributions and interactions with the ISM).

Therefore, we consider the generalized hydrodynamical equation with a vanishing potential: \begin{equation}
{\mu} \, (\frac{\partial}{\partial t} + V \cdot \nabla) V  = -\nabla (Q), \end{equation} The velocity seems
to be invariant during the ejection process. To simplify the search of stationary solutions, it is easier to
rewrite the Euler-Newton/continuity system (Eq. \ref{AA1}, Eq. \ref{AA2} depending on the variables
$[\rho,V]$) in terms of a unique complex equation (depending on the variables $[\textbf{r},t]$) since it
takes a Schr\"odinger-like form, for which several analytical solutions are known. This equation emerges
from the integration of Eq.\ref{schroderiv} and writes (\cite{liwos}, \cite{dirac}) \begin{equation}
{\cal D}^{2}\lapla f + i {\cal D}\frac{ \partial f}{\partial t} - \frac {\Phi}{2{\mu}} f = 0. \end{equation}
In a constant potential, stationary solutions of the form $f (\textbf{r},t)  = 
g(\textbf{r})\,exp({-iEt}/{2{\mu} \cal D})$, can be searched, and we obtain: \begin{equation} 2{\mu}  {\cal
D}^{2} \lapla g(\textbf{r}) -  {E} g(\textbf{r}) = 0, \end{equation} where $E =p^{2} /{2{\mu}} =2{\mu} 
{\cal D}^{2} k^{2}$ is the energy of this free particle. If we take a constant potential different from
zero, we obtain the same equation but with a different particle momentum. The property of the Laplacian in
the spherical coordinates governs the use of the spherical harmonics (which are common functions of $L^2$
and $L_{z}$). The general solution is now constrained by the conservation of the angular momentum and we can
take a solution of the form $g(\textbf{r})  = R(r)\,Y_{l}^{m}(\theta,\phi)$, where $l$ and $m$ are integers.
The radial part is solution of the equation \begin{equation} R^{''}(r) \,+\, \frac{2}{r}
R^{'}(r)\,+\,\left[k^{2} -\frac{l(l+1)}{r^{2}}\right] R(r) = 0. \end{equation} Two radial solutions of this
free particle equation are known \cite{landau}:  \begin{equation} \label{sphericalwave} R_{kl}^{\pm}(r) \,=
\,(-1)^{l} A \frac{r^{l} }{ k^{l}} \left( \frac{1}{r} \frac{d}{dr} \right)^{l} \,\frac {e^{\pm ikr}}{r}.
\end{equation} These functions can be developed in terms of the first order Hankel functions, namely
\begin{equation} R_{kl}^{\pm}(r) \,=\, \pm iA \sqrt{\frac{k \pi}{2r}}  H_{l+\frac{1}{2}}^{(1,2)}(kr),
\end{equation} which represent divergent spherical waves (+ case, with first order Hankel function) or
convergent spherical waves ($-$ case, with second order Hankel function). Divergent spherical waves can
correspond to a flow of central particles emission and the convergent one to an infall flow of particles (we
can already note that the physical solutions are totally reversible: this particular behavior will be
developed in a forthcoming section). The normalization condition restricts the value of the constant, so
that $A={1}/{\sqrt{2{\cal D} k}}$. The square of the modulus of the function $f$ gives the probability
density: \begin{equation} |g(\textbf r)|^{2}\, =\, |R_{kl}(r)|^{2}  |Y_{l}^{m}(\theta,\phi)|^{2}.
\end{equation} We expect the true matter distribution to follow preferentially the peaks of this probability
density distribution. 

Now, this equation represents the distribution of potential trajectories for particles ejected during a unit
time. But our aim is to know the evolution of these trajectories for distances and times higher than the
ejection area. As a first step, the emitted particle flow can be considered to be only radial. Thus, the
progression of the flow is free along the radial coordinate, and only the radial part of the function
depends on the time variable. Thus the progression of the maximum probability can be approximately described
as a spherical wave front with a constant velocity $V_{0}$. The time dependent function simply reads:
\begin{equation} g(\textbf{r},t) \,=\, \frac  {1}{r^{2}} R_{kl}(r-V_{0} t)   Y_{l}^{m}(\theta,\phi).
\end{equation} The $1/r^{2}$ coefficient accounts for the dilatation of the elementary shell.

\subsection{Effect of perturbations} Let us now go on with our progressive method and include, for a more
complete description, the perturbative potential terms. 

First, one can relax the hypothesis of constant velocity by the consideration of a power series expansion of
the radial velocity. These radial perturbative terms only affect the radial part of the spherical potential.
Then this restriction will only change the radial part of the solution, but not the angular part. The
information about the matter density distribution along the shells is related to the ejection history, while
the global shape (which is observed in the end) is given by the angular distribution of matter
(theoretically described in our framework by spherical harmonics). Thus, the principal actors of the
structuration are the symmetries of the system. We can generalize this argument by including all kind of
radial perturbative potentials (compatible with ejection processes) without any change in the result
obtained about morphogenesis.

Other perturbations could affect the global process.  We describe, hereafter, the effect of two of them:\\
$\bullet$ \textit{Self gravitation:} The test-particle ejection associated with bipolar solutions (initially
associated with bipolar axisymmetrical spherical harmonics) evolves through a new dynamical perturbation
field linked to the axial symmetry. A self gravitational potential is created by the initial cylindrical
repartition: each particle is subjected to a perturbative force in $-1/ \rho^2$ (where $\rho$ is the radial
cylindrical coordinate). This attractive gravitational force constricts the ejection cone toward the
symmetry axis of the system. \\ $\bullet$ \textit{Magnetic field:} A magnetic contribution must be
introduced in terms of a perturbative field. Strong magnetic fields are naturally associated with several
stellar stages (e.g. red-giant stars and white dwarf stars \cite{Gseg}). A poloidal field can be
introduced, as a first approximation, and the evolution of ionized particles (because of the UV radiation)
is described by the same dynamics: a constriction along the axial system symmetry.

\section{Results and comparisons to the observational data}

\subsection{Theoretical results}

The main result obtained in the present work is the expected discretization of the possible shapes and the
quantization of conservative quantities associated to the various predicted morphologies.

Let us first consider the main conservative quantity, i.e., energy. The energy to mass ratio ($E/{\mu} = 2 {\cal
D}^2 k^2$) is not quantized in this case, since there is no boundary condition to restrict the value of the
particle wave vector $(k)$. However, one will be led in future developments to consider the connection
between the initial ejection process and the internal star structure. This work could provide a restriction
of the wave vector possible values.

The second and the third prime integrals (consistent with the spherical coordinates) involve a quantization
of the square of the reduced angular momentum and of its projection on the $z$ axis, namely, 
$(L/{\mu})^2=2l(l+1){\cal D}^2$ and $L_{z}/{\mu} =2 m \cal D$, which are independent of the test mass. The ratio
$L/L_z=\sqrt{l(l+1)}/m$ is also independent of the scale-parameter $\cal D$. Thus, this relation gives a
general description of the angular momentum conservation law in all spherical ejection processes,
independently of the specificity of the underlying fractal process.

The interpretation of these solutions is that the matter is expected to preferentially fill the high
probability regions, as defined by the geometry of the geodesics (provided there is a sufficient quantity of
ejected matter). Thus, one can directly associate a quantized  structure with the angular functions.
Moreover, as we shall now see, spherical harmonics involve also a quantization of other observables (i.e.,
the existence of peaks in their probability density distribution). 

The $\theta$ angle is also subjected to a probability density distribution that have quantized peaks. This
angle is observable in terms of the initial ejection angle (in the nearest central area, where the
perturbations are insignificant). Concerning the remaining structure, the theoretically predicted value of
the angle is no longer observable since the perturbation effects (studied above) result in a deformation of
the shape toward the symmetry axis of the system. Another consequence of using spherical harmonics for the
morphology description is the dependence of the matter density distribution on the angle quantization. For
each component of the system (i.e. axial ejections, bipolar shells or equatorial discs), one can predict an
initial ejection angle probability and a relative matter density distribution (relative to the other
components of the matter distribution). All these results about the expected more probable values of the
observable variables are summarized in Table \ref{result}: For each couple $(l,m)$, we give the most
probable values of the ratios $L^2/{\mu}$, $L_z/{\mu}$ and we plot the associated angular solutions, $\rho (\theta)=
{\left| Y_l^{m} (\theta,cst) \right|}^2$ (in a simplified form that takes account of the symmetry under the
$\phi$ rotation). 

Another important information in this table concerns the ratios of the amplitudes of the probability peaks
for the ejection angles. The relative matter density between the imbricated conic structures (see Table
\ref{morpho}) is expected to be proportional to the relative probability density: such a theoretical
expectation can be put to the test (by an analysis of the luminosity ratios) in the observational data (see
example hereafter).

\begin{table*} \centering \rotatebox{90}{ \begin{tabular}{|c|c|c|c|c|c|c|c|c|} 
&$m$=0&$m$=1&$m$=2&$m$=3&$m$=4&$m$=5&$m$=6&$m$=7\\ $\rho (\theta)= {\left| Y_l^{m} (\theta,cst) \right|}^2 $
& & & & & & & & \\ &$L_z /{\mu} =0$&$L_z /{\mu} =2 \,\cal D$&$L_z /{\mu} =4 \,\cal D$&$L_z /{\mu} =6 \,\cal D$&$L_z /{\mu}
=8 \,\cal D$&$L_z /{\mu} =10 \,\cal D$&$L_z /{\mu} =12 \,\cal D$&$L_z /{\mu} =14 \,\cal D$\\ \hline \hline $l$=0&\\
$(L/{\mu})^2 = 0$&\includegraphics[width=2.0cm]{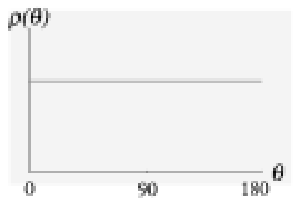} \\ \hline $l$=1&&\\ $(L/{\mu})^2 =4 \,{\cal
D}^2$&\includegraphics[width=2.0cm]{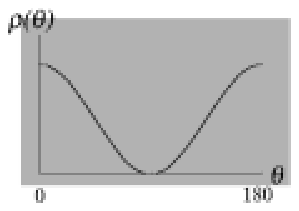}&\includegraphics[width=2.0cm]{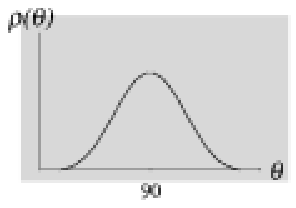} \\
\hline $l$=2&&&\\ $(L/{\mu})^2 = 6 \,{\cal
D}^2$&\includegraphics[width=2.0cm]{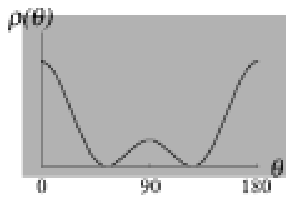}&\includegraphics[width=2.0cm]{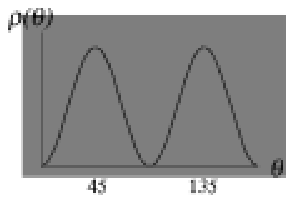}&\includegraphics[width=2.0cm]{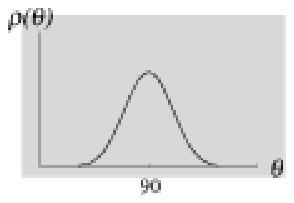}
\\ \hline $l$=3&&&&\\ $(L/{\mu})^2 =8 \,{\cal
D}^2$&\includegraphics[width=2.0cm]{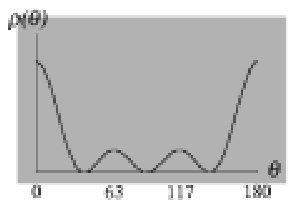}&\includegraphics[width=2.0cm]{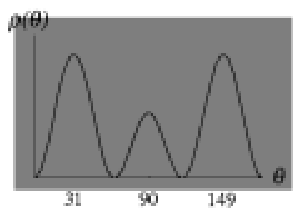}&\includegraphics[width=2.0cm]{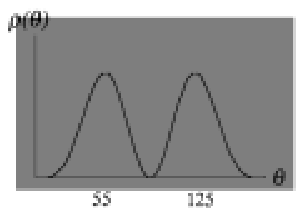}&\includegraphics[width=2.0cm]{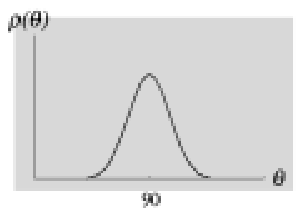}
\\ \hline $l$=4&&&&&\\ $(L/{\mu})^2 = 10 \,{\cal
D}^2$&\includegraphics[width=2.0cm]{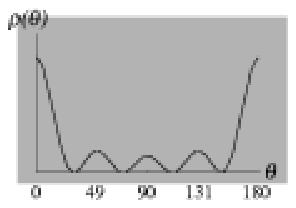}&\includegraphics[width=2.0cm]{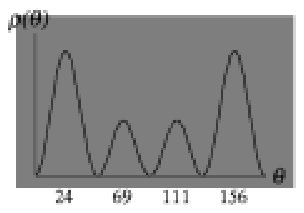}&\includegraphics[width=2.0cm]{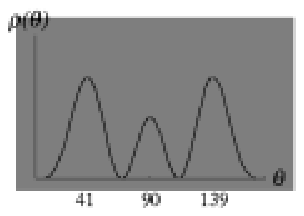}&\includegraphics[width=2.0cm]{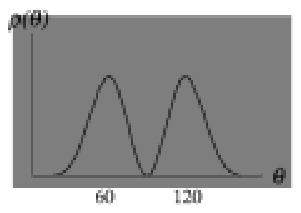}&\includegraphics[width=2.0cm]{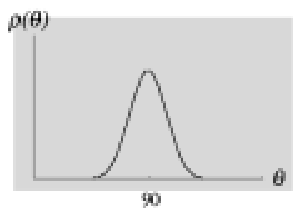}
\\ \hline $l$=5&&&&&&\\ $(L/{\mu})^2 =12 \,{\cal
D}^2$&\includegraphics[width=2.0cm]{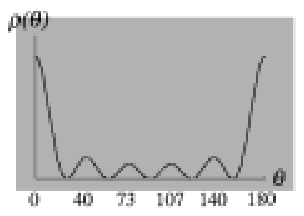}&\includegraphics[width=2.0cm]{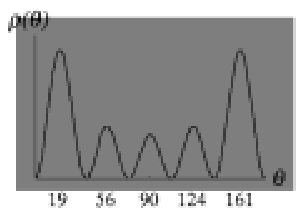}&\includegraphics[width=2.0cm]{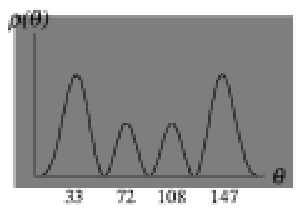}&\includegraphics[width=2.0cm]{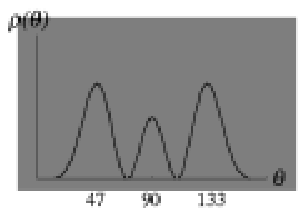}&\includegraphics[width=2.0cm]{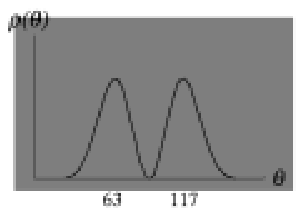}&\includegraphics[width=2.0cm]{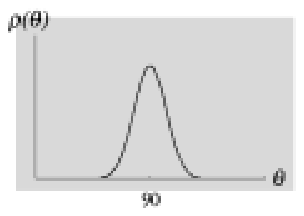}
\\ \hline $l$=6&&&&&&&\\ $(L/{\mu})^2 = 14 \,{\cal
D}^2$&\includegraphics[width=2.0cm]{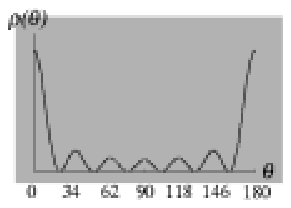}&\includegraphics[width=2.0cm]{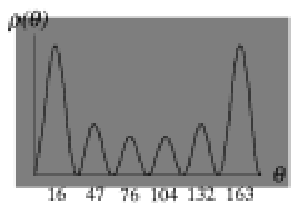}&\includegraphics[width=2.0cm]{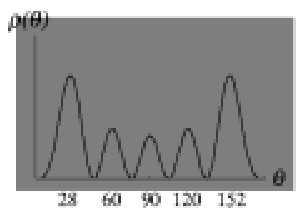}&\includegraphics[width=2.0cm]{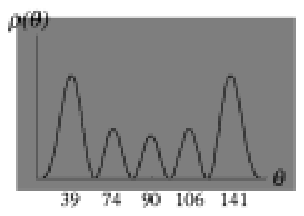}&\includegraphics[width=2.0cm]{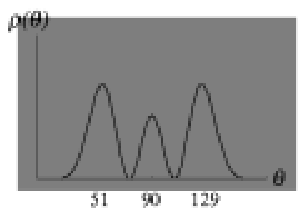}&\includegraphics[width=2.0cm]{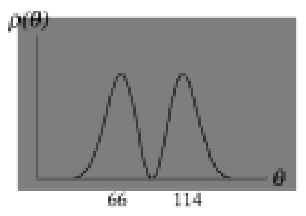}&\includegraphics[width=2.0cm]{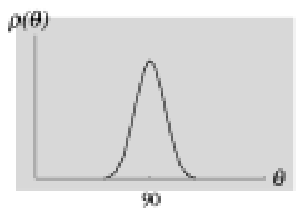}
\\ \hline $l$=7&&&&&&&&\\ $(L/{\mu})^2 =16 \,{\cal
D}^2$&\includegraphics[width=2.0cm]{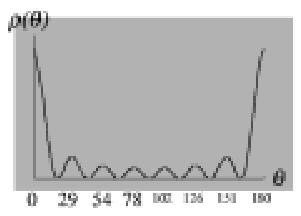}&\includegraphics[width=2.0cm]{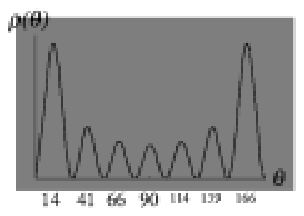}&\includegraphics[width=2.0cm]{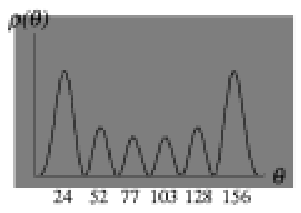}&\includegraphics[width=2.0cm]{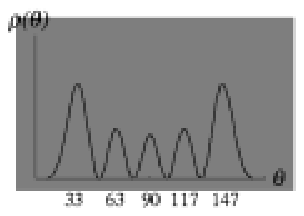}&\includegraphics[width=2.0cm]{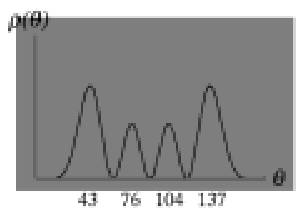}&\includegraphics[width=2.0cm]{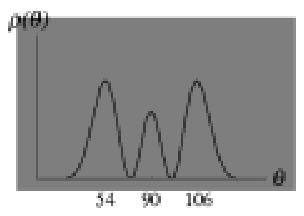}&\includegraphics[width=2.0cm]{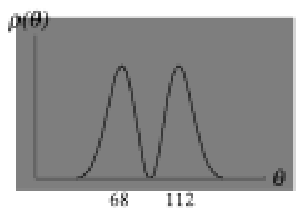}&\includegraphics[width=2.0cm]{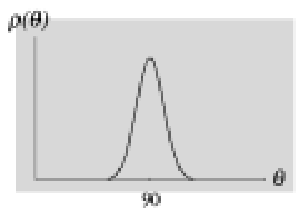}
\\ \hline \end{tabular} } \caption{\footnotesize{Synthesis of the quantized values of the squared angular momentum $L^2/{\mu}$, of its projection along $z$ axis $L_z /{\mu}$, and of the matter distribution along quantized angles $\rho(\theta)$, associated with the angular wave functions $Y_l^{m}$, for the $l$ and $m=1$ to $7$.}}
\label{result} \end{table*}

The various shapes allowed by the generalized hydrodynamical solutions are presented in Table \ref{morpho}.
The bipolar shells are plotted with a weak deviation toward the axis of symmetry. This deviation is typical
of the various perturbative effects considered here-above (e.g., self gravitation, magnetic field).

\begin{table*} \centering \rotatebox{90}{ \begin{tabular}{|c|c|c|c|c|c|c|c|} \hline
&$m$=0&$m$=1&$m$=2&$m$=3&$m$=4&$m$=5&$m$=6\\ \hline
l=0&\includegraphics[angle=-90,width=1.3cm]{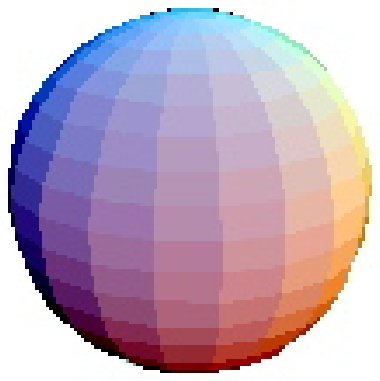}\\ \hline
l=1&\includegraphics[angle=-90,width=2.8cm]{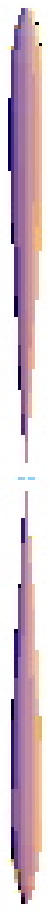}&\includegraphics[angle=180,width=2.5cm]{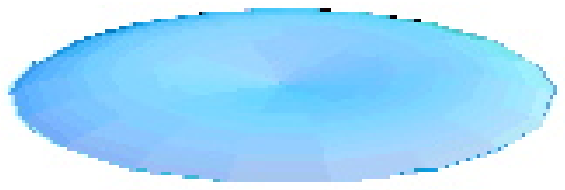}
\\ \hline
$l$=2&\includegraphics[angle=-90,width=2.8cm]{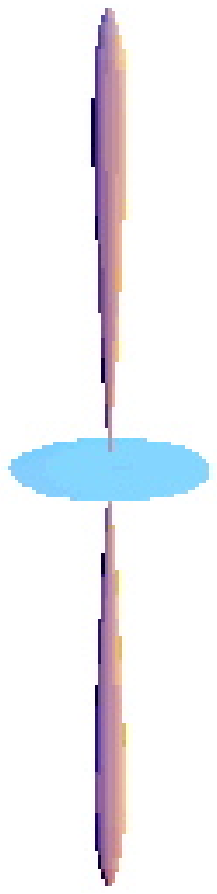}&\includegraphics[angle=-90,width=2.1cm]{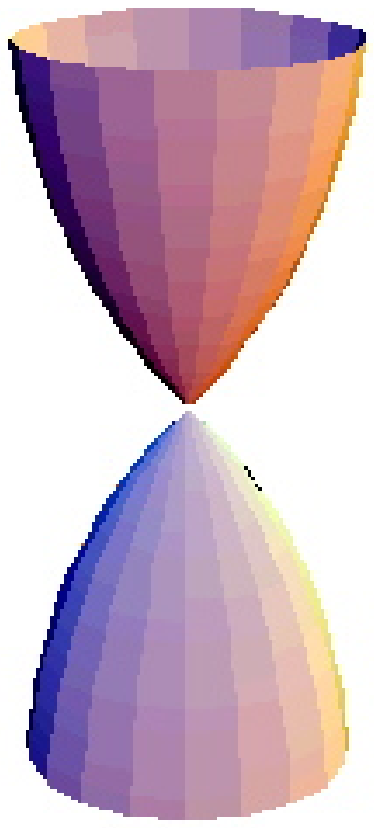}&\includegraphics[angle=180,width=2.5cm]{images/l=m.eps}
\\ \hline
$l$=3&\includegraphics[angle=-90,width=2.8cm]{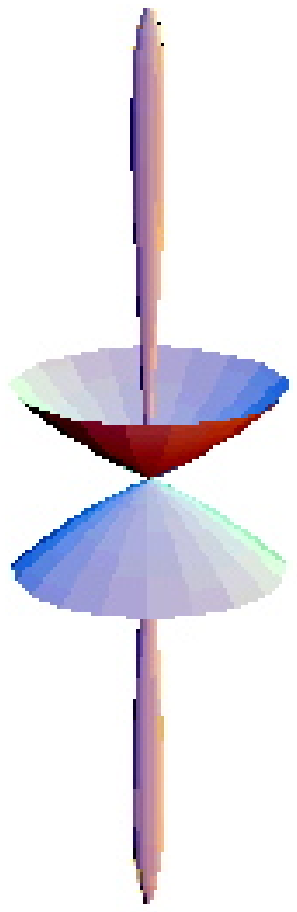}&\includegraphics[angle=-90,width=2.3cm]{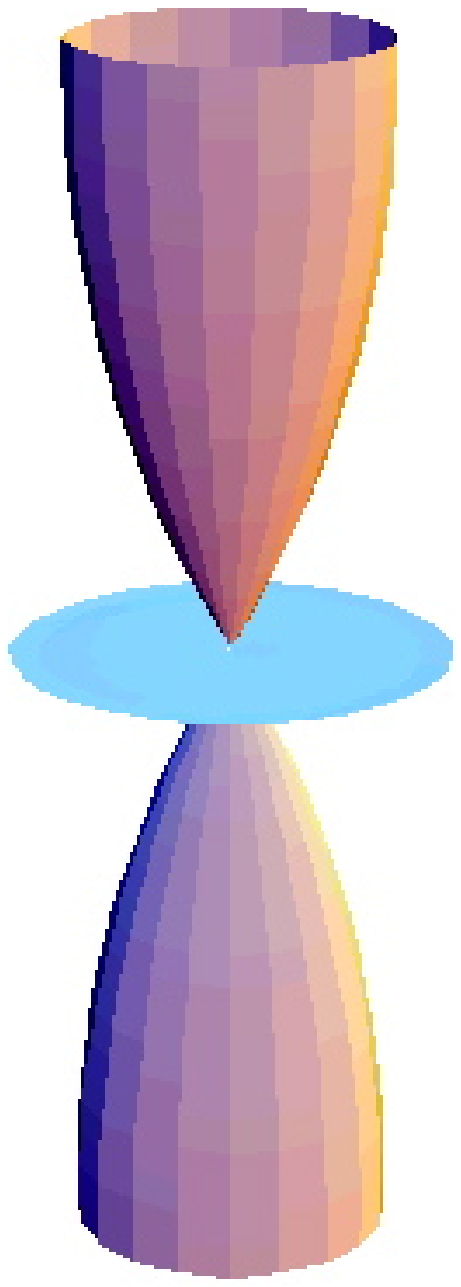}&\includegraphics[angle=-90,width=2.1cm]{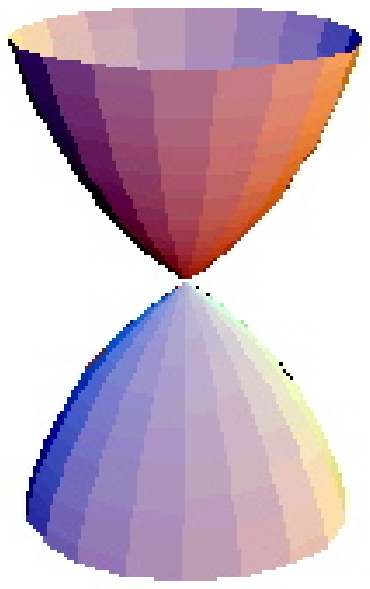}&\includegraphics[angle=180,width=2.5cm]{images/l=m.eps}
\\ \hline
$l$=4&\includegraphics[angle=-90,width=2.8cm]{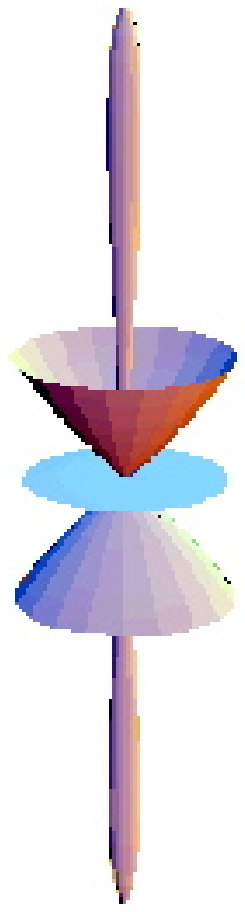}&\includegraphics[angle=-90,width=2.5cm]{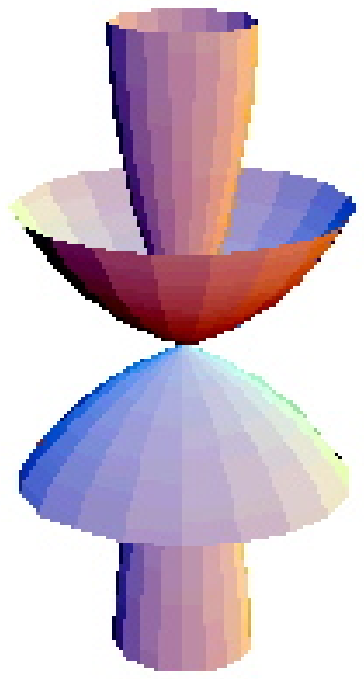}&\includegraphics[angle=-90,width=2.3cm]{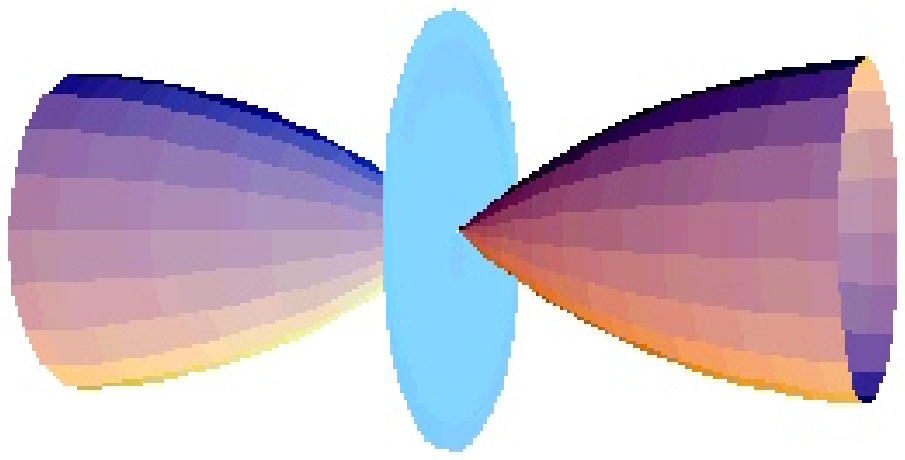}&\includegraphics[angle=-90,width=2.1cm]{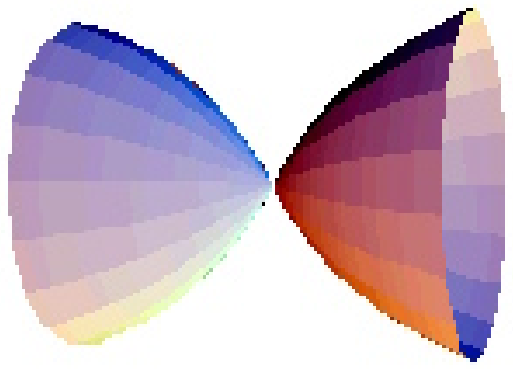}&\includegraphics[angle=180,width=2.5cm]{images/l=m.eps}
\\ \hline
$l$=5&\includegraphics[angle=-90,width=2.8cm]{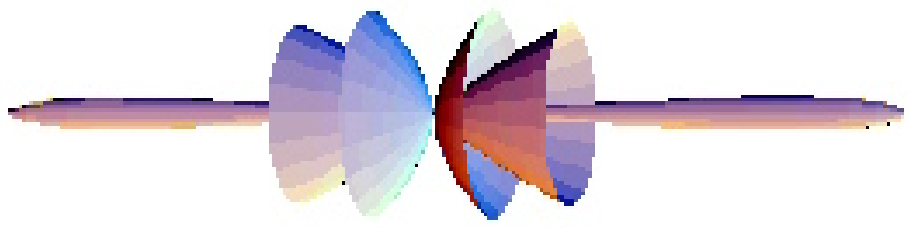}&\includegraphics[angle=-90,width=2.5cm]{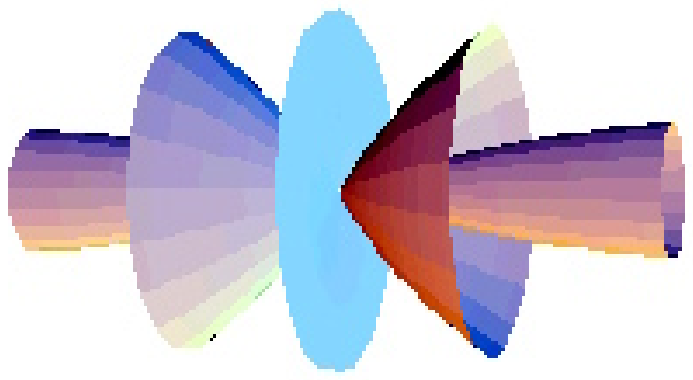}&\includegraphics[angle=-90,width=2.5cm]{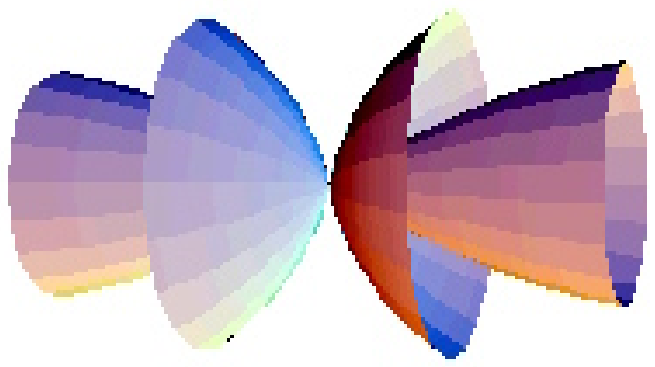}&\includegraphics[angle=-90,width=2.3cm]{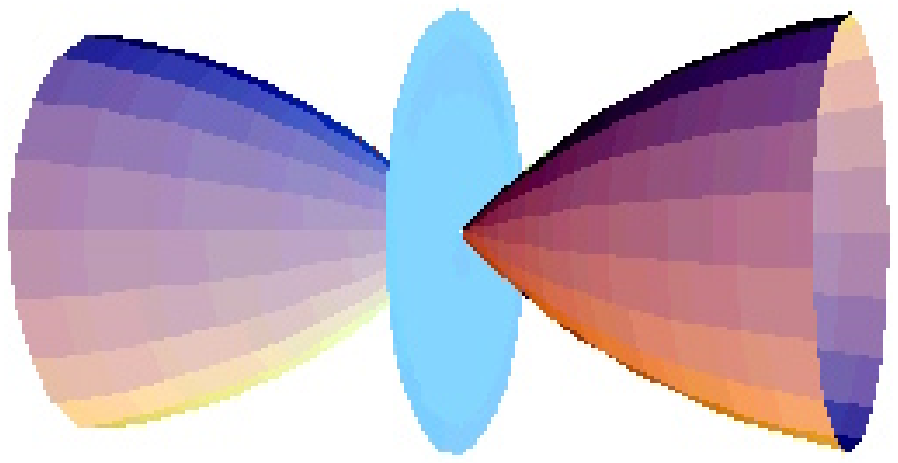}&\includegraphics[angle=-90,width=2.1cm]{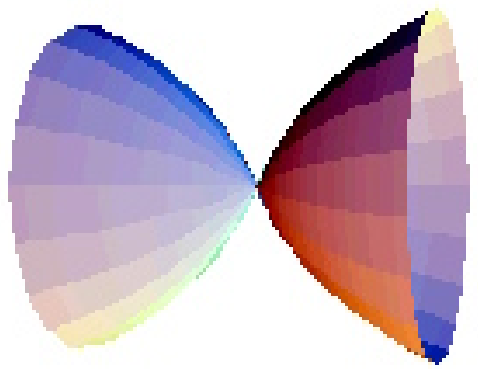}&\includegraphics[angle=180,width=2.5cm]{images/l=m.eps}
\\ \hline
$l$=6&\includegraphics[angle=-90,width=2.8cm]{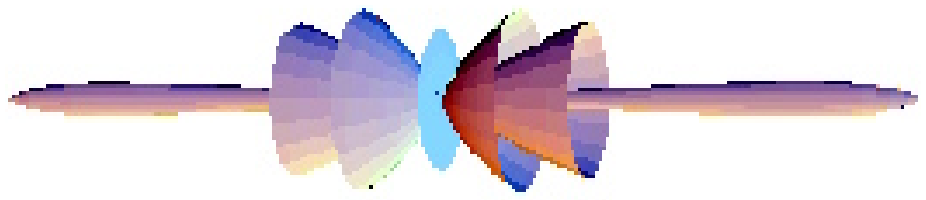}&\includegraphics[angle=-90,width=2.5cm]{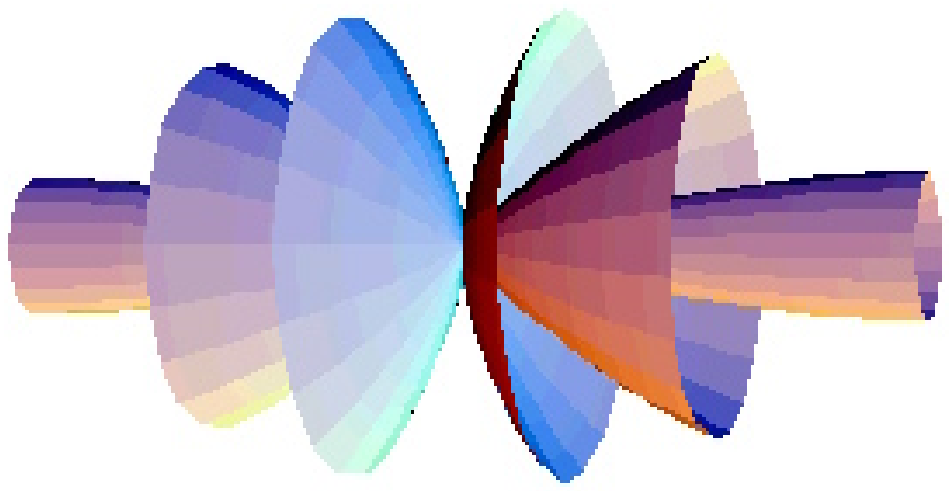}&\includegraphics[angle=-90,width=2.5cm]{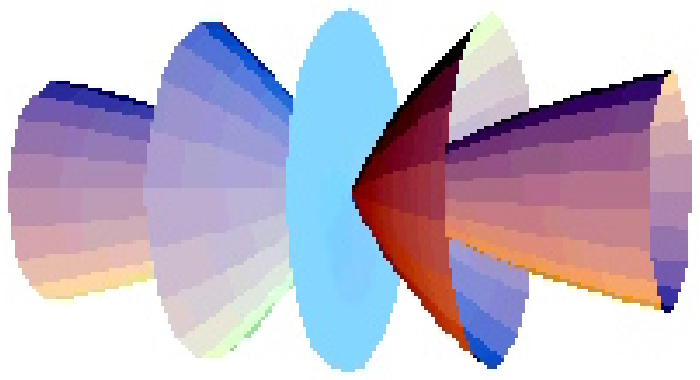}&\includegraphics[angle=-90,width=2.5cm]{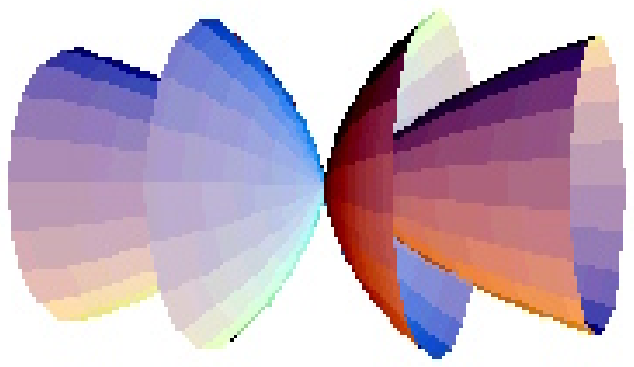}&\includegraphics[angle=-90,width=2.3cm]{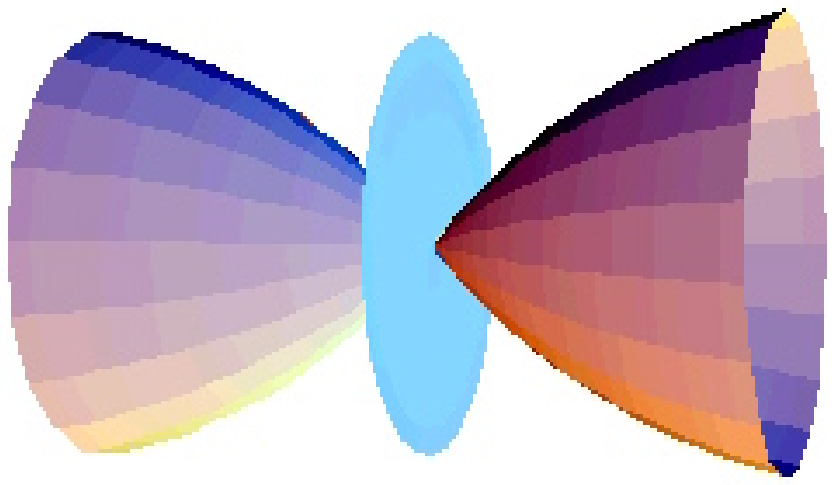}&\includegraphics[angle=-90,width=2.1cm]{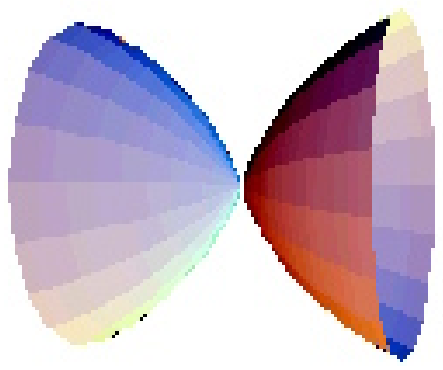}&\includegraphics[angle=180,width=2.5cm]{images/l=m.eps}
\\ \hline
$l$=7&\includegraphics[angle=-90,width=2.8cm]{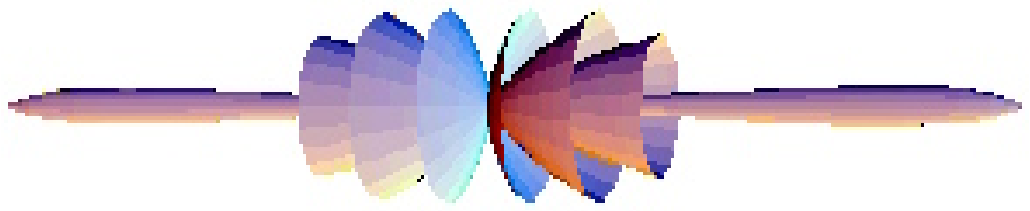}&\includegraphics[angle=-90,width=2.5cm]{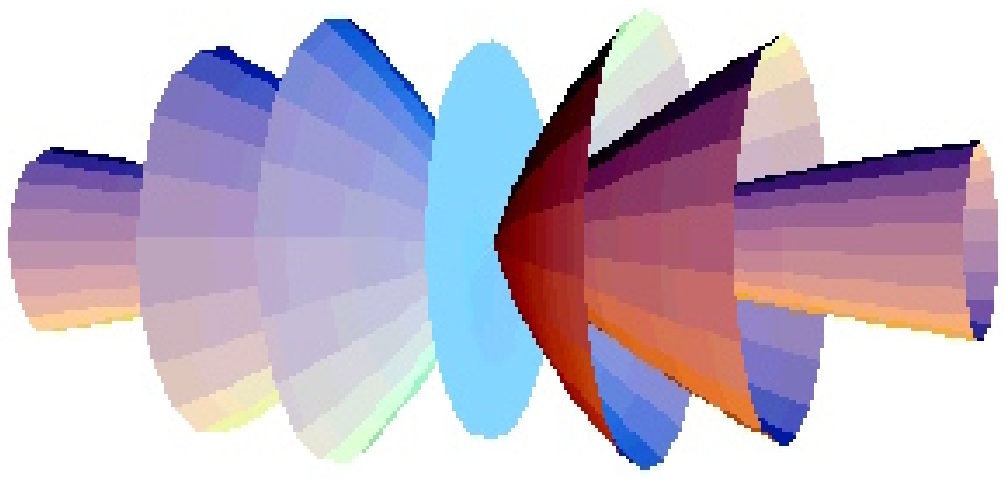}&\includegraphics[angle=-90,width=2.5cm]{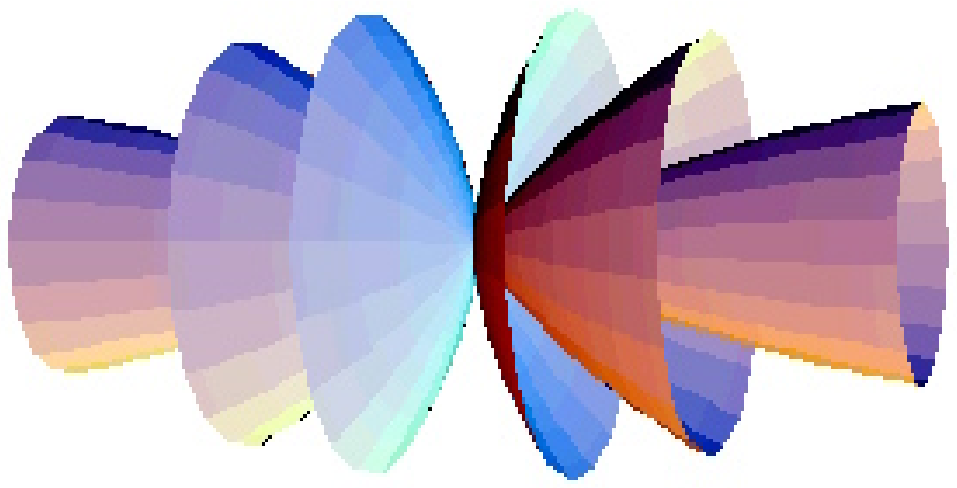}&\includegraphics[angle=-90,width=2.5cm]{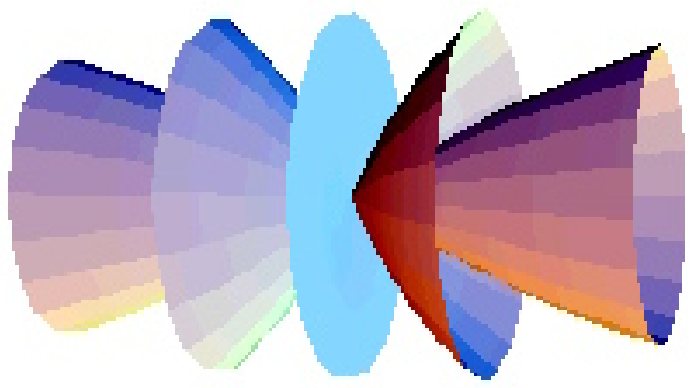}&\includegraphics[angle=-90,width=2.5cm]{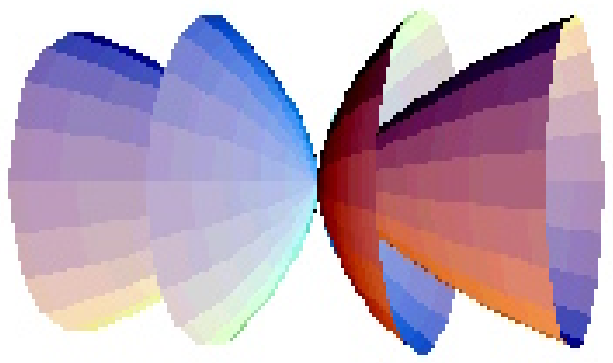}&\includegraphics[angle=-90,width=2.3cm]{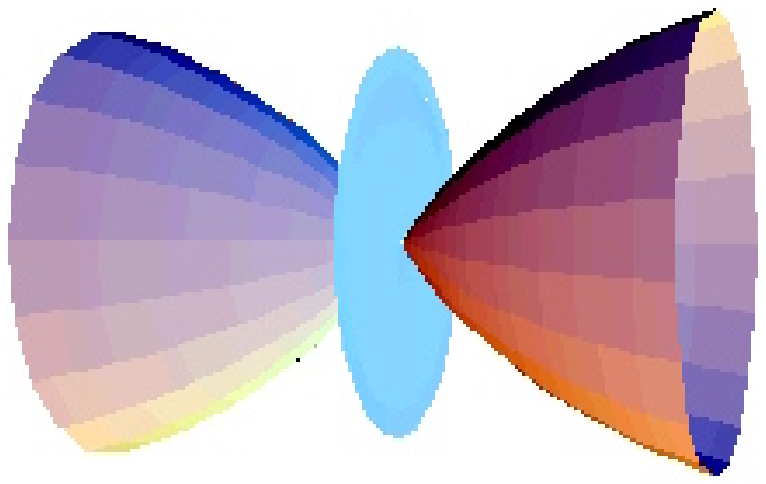}&\includegraphics[angle=-90,width=2.1cm]{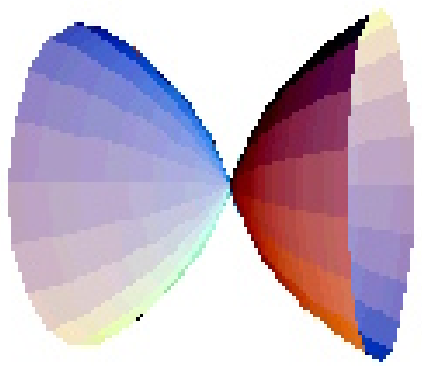}\\
\hline \end{tabular} } 
\caption{\footnotesize{Synthetic representation of the first quantized shapes resulting from the angular ejections solutions given in (Table \ref{result}). Note the analogy of structure of the shapes along diagonals, i.e. $l-m=k=$cst, which are characterized by the same number of imbricated cones, and differ only by the values of the angles. Note also that the theoretically predicted density significantly varies between the cones (see the corresponding figures in Table \ref{result}).}}
\label{morpho}
\end{table*}

\subsection{Comparison with observational data}

\subsubsection{Planetary nebulae} As recalled above, the allowed shapes only depend on the $(l,m)$ couple.
Thus, the expected morphologies are no longer a consequence of the only initial conditions; they are mainly
linked to the values of the conservative quantities and related to the angular boundary conditions. Similar
basic hypotheses allow one to understand the constitution of bright shells in the spherical, elliptic and
bipolar planetary nebulae.   For remaining uncommon morphologies, various perturbations could be included in
the active potentials. Moreover, more complicated symmetries (linked to other dynamical conservation laws)
remain to be explored.\\ \textbf{A new morphological description of planetary nebulae:} \\ $\bullet$
\textit{Spherical}: \begin{figure}  \centering \includegraphics[width=6cm]{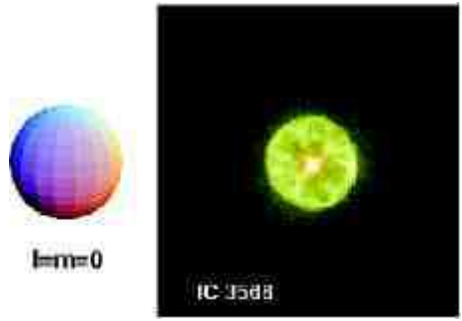} \caption{The
elementary isotropic structure emerges from the fundamental level of the angular quantization law
$(l=0,m=0)$. Left hand side: theoretical prediction. Right hand side: the spherical planetary nebula IC
3568. } \label{fig:sper} \end{figure} For a $(l,m)$ couple equal to $(0,0)$, one expects an isotropic matter
distribution. This elementary case reflects the properties of the spherical planetary nebulae which were
first discovered.\\ $\bullet$ \textit{Elliptic}: All the solutions consistent with $(l=m)$ involve an
equatorial disc structuration. With different tilting on the line of sight, elliptic shapes emerge from this
solution. It is important to note that elliptic shapes can also result from perturbed spherical shapes, in
accordance with the ISW model for elliptic shapes. Moreover, provided some apparently elliptic shapes are
indeed the result of such a projection effect of disk structures, one expects to observe one day extremely
flattened elliptical PNe: however, up to now this kind of PNe have not yet been observed (to our knowledge).
\begin{figure}  \centering \includegraphics[width=8cm]{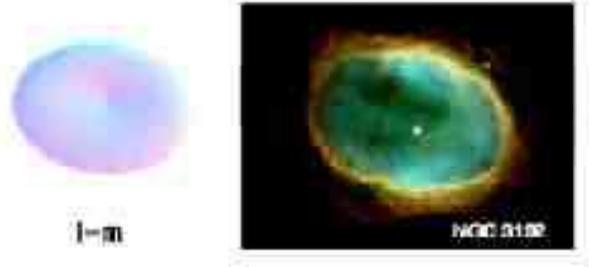} \caption{For some values of the
inclination, the theoretical equatorial disc (left hand side) could appear like an elliptic PN. For example,
NGC 3132 is observed elliptic, but there is still discussions about its true morphology (Monteiro et al.
2000).} \label{fig:elliptic} \end{figure}\\ $\bullet$ \textit{Bipolar}: Bipolar shapes emerge naturally from
this work. These shapes are, in the standard approach, explained in terms of various effects such as
different density outflows, presence of a double star system, etc... The present approach does not
contradict these views: we simply connect the shapes directly to the state of the system, as described in
terms of its conservative quantities (at first, the state of angular momentum). Whatever the particular
cause for the system to jump to this state (which can be the presence of a second star, etc...), the same
shape is expected to be observed for a given state.  Moreover, this result generalizes the ISW model
principles that lead to bipolar structures, since the free particles in bright shells  could be directly
issued from wind shocks (which are not conserved in standard models). 

\begin{figure}  \centering
\includegraphics[width=8cm]{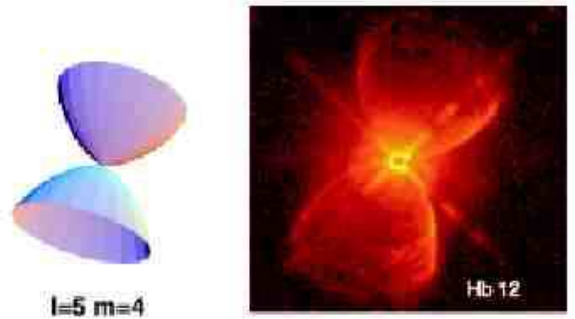} \caption{Elementary bipolar structure: the simple shell
structure. The predicted ($5,4$) quantization with an initial angle of $63^\circ$ is consistent with the
observed  structure of the planetary nebula $Hb12$ .} \label{fig:bipo1} \end{figure}

The first bipolar structure allowed by the possible values of the angular momentum is the simple shell one.
This kind of simple situation can be illustrated by the planetary nebula $Hb12$ (fig. \ref{fig:bipo1}). The 
initial angle is consistent with a ($l=5, m=4$) quantization (near $63^{\circ}$) but this  value is not
sufficiently well measured to specify the quantum numbers with certainty (with regard to the nearest
possibilities: ($l=4, m=3$) or ($l=6, m=5$)).  \\ Bipolar shapes become
more complicated when the angular momentum increases. Beyond simple shells ($m=l-1$), bipolar shells with
equatorial disks emerge in the same way ($m=l-2$, cf the$Hb 5$ structure). 

\begin{figure}  \centering \includegraphics[width=7cm]{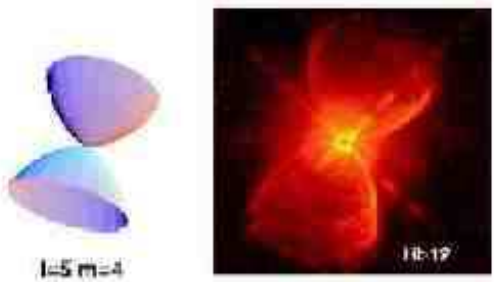} \caption{Double shell structure:
the M2-9 planetary nebula can be associated with an angular momentum quantization of quantum numbers
($l=5,m=2$).} \label{fig:bipo2} \end{figure}

The next case (fig. \ref{fig:bipo2}) is the double-shell planetary nebula of the M2-9 type. This shape is
naturally expected from our classification, as corresponding to $m=l-3$ (table \ref{morpho}). The morphology
of M2-9 can be associated to the quantum numbers ($5,2$) (cf. table \ref{result}) which involves a quantized
angular distribution around the values $(33^{\circ},72^{\circ},108^{\circ},147^{\circ})$ and a higher density 
for the inner shells. The PN structure near the central star is not clearly defined, but the extrapolation of the
external shell allows us to estimate the  values of the initial ejection angles. This estimation ($\approx
30^\circ$ for the outer shell and $\approx 70^\circ$ for inner one) is consistent with the predicted angles:
moreover,  the inner observed density is clearly higher than the outer shell density, as theoretically
expected.\\ Even more complex (therefore less probable) structures that have not been observed up to now,
are still possible: double-shell + disk ($m=l-4$), triple-shell ($m=l-5$), etc... In all these new cases, as
for the already observed ones, the theoretical prediction applies not only to the overall morphology, but
also to quantitative values of angles, angular momenta and density distributions, and it will therefore be
possible to put it to the test in the near future.

\subsubsection{LBV stars} The theoretical development presented in the first part of this paper is
associated with a general ejection process. Luminous blue variable (LBV) stars are in essence very different
from PN central stars. Despite of this internal incompatibility, some recent works \cite{dwarbal} have been
developed to explain the shell formation around LBV stars by the same hydrodynamical ISW model that was used
for planetary nebulae. 

\begin{figure}  \centering \includegraphics[width=8cm]{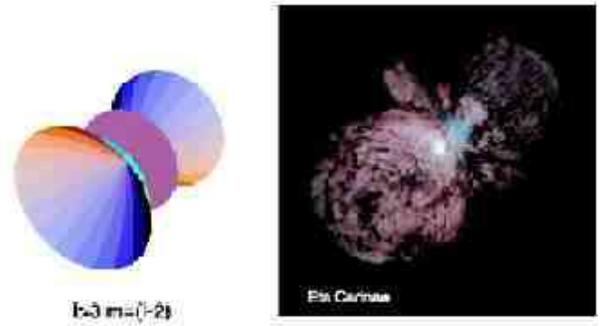}
\caption{Generalization to another singular stellar systems: comparison between theoretical shapes with
simple bipolar shell/equatorial disc and a LBV star (Eta Carinae).} \label{fig:eta} \end{figure}

Therefore the method studied in the present paper can be applied to a morphological description of these
objects. In consequence we also expect a quantization of quantities such as angular momenta and angles and a
discretization of the possible shapes for the matter density distribution around LBV stars. For example, the
Eta Carinae LBV star shows bipolar shell with an equatorial disk (fig. \ref{fig:eta}: such a structure is
expected in all the cases when $l>3$ and $m=(l-2)$). Therefore this result supports the idea of a general 
behavior of ejection processes and of the self-similarity of shell structurations, independently of the
nature of the central star.  

\subsubsection{Supernovae} 

There are other systems that can be analyzed with this new point of view, in particular supernova remnants.
Whatever the specific ejection process at work in supernovae (Ia or II..), the three conditions upon which
our method relies, namely (i) fractal trajectories, (ii) infinity of potential trajectories, (iii)
time-reflection breaking, are expected to be achieved in this case, since the same collisional/diffusing
processes as for planetary nebulae are present. Moreover, the hereabove theoretical developments can be
generalized to scattering systems described by an initial convergent wave function (as in implosion
processes) and by a resultant divergent wave function (ejection processes). Supernova structures emerge from
the interaction of remnant matter from the central star explosion and the ISM. This simple presentation
suggests a behavior of SNe particles which may be similar to ejected particles in PNe or LBV stars. Thus,
provided the ISM interaction is assumed to be homogenous (as a first order approximation), the quantized
results obtained by the theoretical development (table \ref{morpho}) are still valid. The account of
inhomogeneities of the ISM would be more difficult, since they degrade the initial conserved information on
prime integrals and then, would involve a more complicated description. \begin{figure}  \centering
\includegraphics[width=8cm]{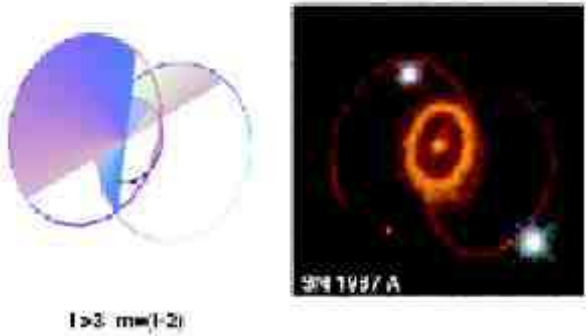} \caption{Simple model of supernova remnant: as a first order
approximation, one finds that supernovae remnants could also be structured like other spherical ejection
systems. For a single ejection phase, this leads to the formation of expanding rings, that are expected to
be ejected with higher probability along given angles.} \label{fig:SN} \end{figure} One of the most
enigmatic SN is SN-1987A (Fig. \ref{fig:SN}). Its particular structure (which has up to now resisted to all
hydrodynamical simulations) can be accounted for by a unique ejection associated with a quantized angular
momentum corresponding to $m=l-2$ (see Table \ref{morpho}). This configuration induces, in the case of a
discontinuous ejection process, a geodesic structuration into two bipolar rings and an equatorial ring, as
observed. Moreover, in recent images the central star is projected on the right ring. In such a
configuration, the ejection cone angle is given by $\cos(\theta)=b/a$ where $a$ and $b$ are respectively the
major and minor axes of the projected ellipse. From 6 measurements on the inner edge, middle and outer edge
of the equatorial ring and of the right polar ring (the left ring was not used because it shows distortions
to ellipticity), we find $\theta=41.2 \pm 1.0$ deg, which supports the identification with the case $m=2,
l=4$, for which the predicted angle is 40.89 deg (while the nearest similar configurations yield 31 deg and
47 deg). The youth of the system (10 years compared with $10^4$ years for typical PNe) supports the
assumption that the observed angle is still the initial ejection angle.

From this preliminary work, we suggest that future observations of young supernovae in homogeneous ISM will
allow one to put the theory to the test in a quantitative way, since we expect the occurrence of quantized
structures consistent with the general results given in Table \ref{morpho}.

\subsubsection{Star formation} Low mass star formation processes seem to be
universally associated with circumstellar infall / accretion discs and outflow / axial ejections
(\cite{andre}, \cite{cabrit}). In the standard approach, the existence of such simultaneous structures
is still not fully understood (despite several attempts using various magnetic models \cite{ferreira}).

The solutions associated with free motion in spherical potentials describes either convergent or divergent
spherical waves (eq. \ref{sphericalwave}). From the point of view of the present theoretical description in
terms of probability amplitudes, the two cases of ejection and accretion both correspond to a scattering
process (one simply reverses the sign of time). It is therefore quite simple to consider solutions that
combine accretion (for example, equatorial) and ejection (fro example, polar).

One cannot make any longer the hypothesis of free motion in an accretion phase. However, this concerns only
the radial dependence of the solutions, while the angular dependence is still expected to be given by
spherical harmonics. Finally, in a typical star birth stage, the above method is able to account for the
observed discretized structuration of the infall matter and of the resultant particle ejection.

\begin{figure}  \centering \includegraphics[width=8cm]{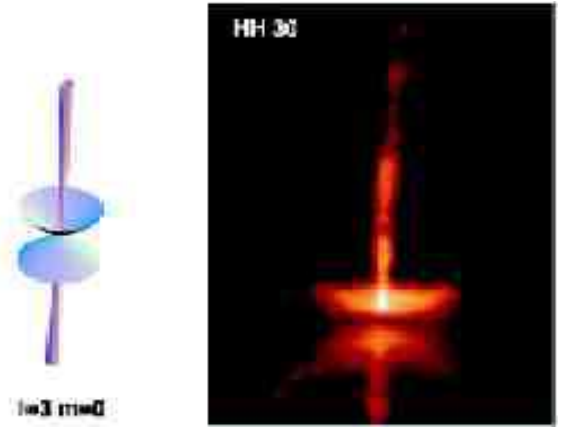} \caption{An example of morphology
observed in a star formation process, compared with a theoretically predicted possible structure.}
\label{fig:hh} \end{figure}

For example the HH-30 observations (Fig. \ref{fig:hh}) allow us to make a preliminary analysis of this
proposal. The equatorial disc and the resultant axial ejection agree with the morphology given by $l=1$,
$m=0$ for axial ejection and $m=l$ for the accretion disc. The global angular momentum is conserved during
the scattering process, but the $L_z$ behavior is still an open problem (since it could be described by a
more complex solution like ($l=2,m=0)$). The bright bipolar shells are currently considered as reflection
nebulae \cite{burrows}, but this component could also have a different dynamical behavior (with respect to
the equatorial disc) and thus could correspond to a solution like ($3,0$) or ($4,0$). We present in Fig.
(\ref{fig:hh}) the structure closest to observations allowed by the theoretical developments.

\section{Conclusion and prospect}

In this paper, we have given a general description of matter structuration in spherically symmetric
potentials. The motion equation, rewritten in terms of an hydrodynamical system which incorporates the
equation of continuity, depends directly on the conditions of symmetry of the system. 

In the case of spherical symmetry, which is the situation mainly considered in this study, the solutions
depend on the values of the conservative quantities ($E$, $L^2$ and $L_z$). This method leads to a great
diversity of stationary solutions, that are expected to describe infall and ejection motion from central
bodies, in terms of probability density distributions for the radial distance and the values of the ejection
(or accretion) angles.

In the case when the potential depends only on the radial coordinate, the general solutions (which give the
density of potential trajectories) are then separable in a radial $R(r)$ and a spherical part given by the
spherical harmonics $Y_l^m(\theta,\phi)$. The possible morphologies are therefore classified according to
the values of the integer numbers $l$ and $m$. Thus, most probable quantized structures (table \ref{morpho})
result from the angular probability distributions (table \ref{result}) and can be compared with the various
systems that come under such a description (planetary nebulae, LBV stars, supernovae, young star formation).
Moreover, the theoretical approach allows one to predict the possible existence of structures that have no
yet been observed, and of yet unidentified correlations between the morphology and observable quantities.

Let us conclude by some examples of possible future extensions of the present study: 

- The connection between internal stellar structures and the ejection process could provide informations
about the probability distribution of the velocity ejection.

 - Concerning the planetary nebula description, the internal stellar angular momentum is known to be
structured: various external quantized solutions are then possible. The internal structuration could induce
an external combination of spherical harmonic solutions (depending on the ejection history ). 

 - Particular quantized solutions ($l,m$) can be connected with the central system and show a dependence on
specific conditions (single star with inclined magnetic fields, double star system etc..). 

 - Several simple systems present axial symmetry: therefore several new possibilities of  morphogenesis can
be explored by considering other coordinate systems (e.g. cylindrical symmetry, parabolic symmetry etc..). 

 - Only hints about the perturbation effects have been given here: future work is needed in order to
introduce the perturbation potential in the Euler-Newton / continuity equation (non trivial solutions). 

 - Numerical simulations using the above solutions as initial conditions can be performed, in order to study
the influence of perturbation fields (e.g. magnetic contribution, self-gravitation, ISM interaction...), 
and  to explore the possibility of appearance of  non-trivial structures.

\section*{Acknowledgments}

We thank Drs T. Lery,  G. Stasinska, D. Pequignot and C. Morisset for helpful discussions.

\end{document}